\documentclass[aip, apl, preprint]{revtex4-1}
\usepackage{graphicx}
\usepackage{dcolumn}% Align table columns on decimal point
\usepackage{bm}% bold math
\usepackage{xr}

\newcommand{\oneoneo}{$\left[110\right]$}

\makeatletter
\newcommand*{\addFileDependency}[1]{% argument=file name and extension
  \typeout{(#1)}
  \@addtofilelist{#1}
  \IfFileExists{#1}{}{\typeout{No file #1.}}
}
\makeatother
 
\newcommand*{\myexternaldocument}[1]{%
    \externaldocument{#1}%
    \addFileDependency{#1.tex}%
    \addFileDependency{#1.aux}%
}

\myexternaldocument{supplement}

\begin{document}

\title{Decoding the complexities of lead-based relaxor ferroelectrics}

\author{Abinash Kumar}
\affiliation{Department of Materials Science and Engineering, Massachusetts Institute of Technology, Cambridge, MA 02139, USA}

\author{Jonathon N. Baker}
\affiliation{Department of Materials Science and Engineering, North Carolina State University, Raleigh, North Carolina 27695-7907, USA}

\author{Preston C. Bowes}
\affiliation{Department of Materials Science and Engineering, North Carolina State University, Raleigh, North Carolina 27695-7907, USA}

\author{Matthew J. Cabral}
\affiliation{Department of Materials Science and Engineering, North Carolina State University, Raleigh, North Carolina 27695-7907, USA}

\author{Shujun Zhang}
\affiliation{Institute for Superconducting and Electronic Materials, Australian Institute of Innovative Materials, University of Wollongong, Wollongong, NSW 2500, Australia}

\author{Elizabeth Dickey}
\affiliation{Department of Materials Science and Engineering, North Carolina State University, Raleigh, North Carolina 27695-7907, USA}

\author{Douglas L. Irving}
\affiliation{Department of Materials Science and Engineering, North Carolina State University, Raleigh, North Carolina 27695-7907, USA}

\author{James M. LeBeau}
\affiliation{Department of Materials Science and Engineering, Massachusetts Institute of Technology, Cambridge, MA 02139, USA}

% \email{jmlebeau@ncsu.edu}

\begin{abstract}

\textbf{Relaxor ferroelectrics, which can exhibit exceptional electromechanical coupling are some of the most important functional materials with applications ranging from ultrasound imaging to actuators and sensors in microelectromechanical devices.  Since their discovery nearly 60 years ago, the complexity of nanoscale chemical and structural heterogeneity in these systems has made understanding the origins of their unique electromechanical properties a seemingly intractable problem. A full accounting of the mechanisms that connect local structure and chemistry with nanoscale fluctuations in polarization has, however, remained a need and a challenge.  Here, we employ aberration-corrected scanning transmission electron microscopy (STEM) to quantify various types of nanoscale heterogeneity and their connection to local polarization in the prototypical relaxor ferroelectric system Pb(Mg$_{1/3}$Nb$_{2/3}$)O$_{3}$-PbTiO3 (PMN-PT). We identify three main contributions that each depend on Ti content: chemical order, oxygen octahedral tilt, and oxygen octahedral distortion. These heterogeneities are found to be spatially correlated with low angle polar domain walls, indicating their role in disrupting long-range polarization.  Specifically, these heterogeneities lead to nanoscale domain formation and the relaxor response. We further locate nanoscale regions of monoclinic distortion that correlate directly with Ti content and the electromechanical performance. Through this approach, the elusive connection between chemical heterogeneity, structural heterogeneity and local polarization is revealed, and the results validate models needed to develop the next generation of relaxor ferroelectric materials.  }
\end{abstract}

\maketitle

Relaxor ferroelectrics are distinguished from traditional ferroelectrics by their frequency dependent, diffuse phase transitions that are commonly attributed to the existence of nanoscale order.\cite{Cohen2006} Among this class of materials, Pb-based systems remain essential as they exhibit the largest known piezoelectric coefficients and outstanding dielectric properties.\cite{Park1997,Zhang2012,Zhang2015}  These properties have been commonly attributed to the existence of polar nanoregions (PNRs) in the material that align to the applied field.\cite{Burns1983,Burns1983b} This simple model fails, however, to explain relaxor behavior in a variety of materials, including polymer-based systems.\cite{Yang2014} The recently proposed ``polar slush'' model overcomes these limitations by considering a multi-domain polar state with low energy, low angle domain walls formed throughout.\cite{Takenaka2017} Model validation has, however, been hindered by the seemingly endless number of structural characterization studies with differing interpretations, leading these systems to be described as a  ``hopeless mess'' .\cite{Cohen2006}

Pb(Mg$_{1/3}$Nb$_{2/3}$)O$_{3}$-PbTiO$_3$  (PMN-PT) is the prototypical Pb-based relaxor ferroelectric system and exhibits state-of-the-art properties.  For example, piezoelectric constants up to 4100 pC/N have been achieved via Sm doping \cite{Li2019}. PMN-PT adopts the perovskite structure (formula ABO$_3$) with the A sub-lattice occupied by Pb and the B sub-lattice occupied by the either Mg, Nb, or Ti. Furthermore, as PbTiO$_3$ is added to PMN, the system reaches a morphotropic phase boundary (MPB) near PMN-30PT where the electromechanical properties are maximized.\cite{Krogstad2018} Deciphering the origins of these enhanced properties is, however, complicated by nanoscale heterogenity of chemistry and structure.  While X-ray and neutron scattering are the predominate characterization methods to explain these details, making direct connections between composition, structure, and polarization have proven exceedingly difficult  \cite{Singh2006,Singh2003,Thomas1999a,Kim2012a,Cowley2011,Davis2007,Randall1990}.  For example, while nanoscale chemical ordering of the B sub-lattice cations was one of the first signatures revealed \cite{Randall1990,C.A.Randall1990,C.A.Randall21990} and is found to varying degrees in all Pb-based relaxor ferroelectrics\cite{TakesueN1999,Goossens2013},  its details are still being revealed  \cite{Cabral2018,Kopecky2016,Eremenko2019}.  

\begin{figure}
    \centering
    \includegraphics[width = 6.4
    in]{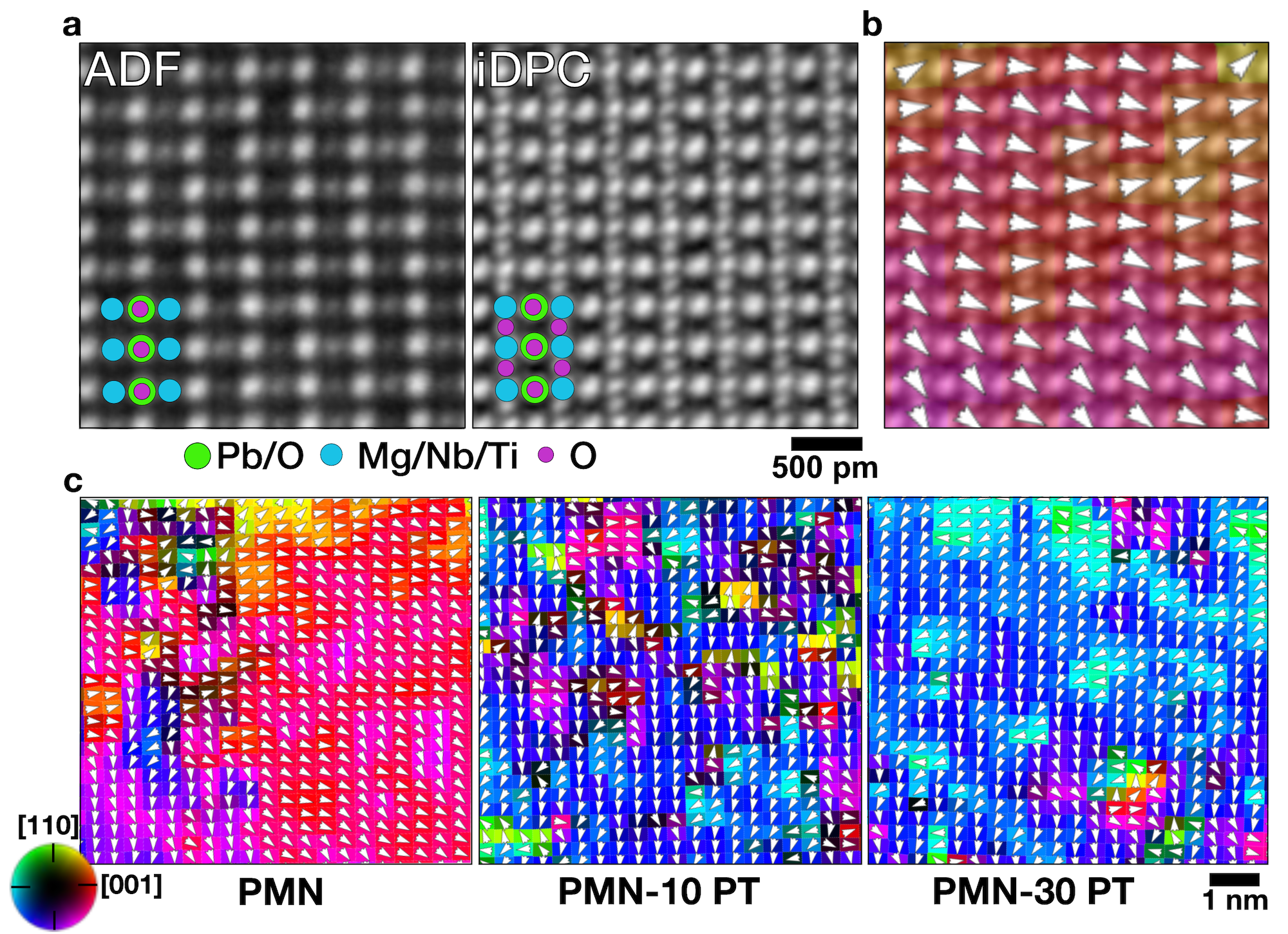}
    \caption{(a) Simultaneously acquired ADF and iDPC images of PMN along the \oneoneo{} projection with the (b) projected polarization map from iDPC. (c) Projected displacement (polarization) for  PMN-xPT (x=0, 10, \& 30) unit cells in representative iDPC images.  The projected displacement(polarization) magnitude ranges from 1 to 20 pm, and is indicated by luminosity.}
    \label{fig:quiver}
\end{figure}

Capturing a complete picture of the connection between chemistry, structure, and polarization in relaxor ferroelectrics requires characterization techniques that are able to account for both cation \textit{and} anion sub-lattices.  While, distortions of the oxygen sub-lattice previously have been  detected with neutron diffraction \cite{Rosenfeld1995}, a combination of X-ray and neutron diffraction  recently discovered that correlated oxygen displacements are likely the key to understanding the electromechanical properties \cite{Krogstad2018}. Diffraction characterization methods, however, lack the spatial resolution required to directly determine the origin of nanoscale polar variation \cite{Keen2015,Xu2008}. In contrast, annular dark-field  scanning transmission electron microscopy (ADF STEM), Figure \ref{fig:quiver}b, can readily image the projected crystal structure at the atomic length scale and is sensitive to chemical ordering \cite{Cabral2018}. These ADF images are, however, dominated by cation contrast that precludes detailed detection and quantification of light elements such as oxygen. Recently, annular bright field (ABF) \cite{Findlay2009,Kim2017} and integrated differential phase contrast (iDPC) \cite{Lazic2016,DeGraaf2018} STEM techniques have been introduced to overcome these challenges.  As shown in Figure \ref{fig:quiver}b, iDPC can visualize the cation and oxygen atom column positions, but lacks direct atomic number interpretability. By simultaneously capturing ADF and iDPC images, however, the resulting datasets are akin to merging results from X-ray and neutron diffraction techniques, and enables the correlation between polarization, chemistry, and structure directly at the atomic scale.

Here, we investigate the structural and chemical origins of  relaxor ferroelectric properties in PMN-xPT (x=0, 10 \& 30). Through a combination of ADF and iDPC aberration corrected STEM, the projected positions of cation and anion sub-lattices are used to measure the subtle features of nanoscale polarization in these materials. The  projected polarization reveals the presence of nanoscale domains that are consistent with a frozen `polar slush' model.\cite{Takenaka2017} We further quantify the distribution of chemical and structural heterogeneities as a function of Ti content, and a direct correlation is found between the spatial distribution of chemical/structural heterogeneities and polarization domain walls.  The heterogeneities are found to inhibit polarization rotation, leading to the formation of low angle domain walls.  In combination, these results provide evidence for the underlying mechanisms responsible for yielding relaxor behavior.

%The of the interpretation is complicated by can be explained by either the existence of PNRs in non-polar matrix or a multi-domain ``polar slush'' state.\cite{Takenaka2017}
%While `butterfly-shaped' diffuse scattering is are associated with relaxor behavior, the addition of PbTiO$_3$ reduces the relaxor behavior in  PMN based relaxor ferroelectrics, while butterfly-shaped diffuse scattering intensity increases with a maxima at the MPB boundary composition. 
%This lead to the conclusion that butterfly shaped diffuse scattering should not be considered as the signature of relaxor system as diffuse scattering intensity does not correlate with the relaxor dielectric behavior.\cite{Krogstad2018}
%Similar shaped diffuse scattering is also observed in non-relaxor system like PZT,\cite{Burkovsky2012a} rendering the interpretation of scattering results is highly ambiguous. 

%%%%%%%%%% Main experimental Results %%%%%%

%  (Pb/O and Nb/Mg/Ti atom)of the  ( and anion (O) atom columns,

%While the Pb/O atom columns contain both Pb and O, the atom column center is largely determined by the Pb displacements as explored in Supplementary Information Section 1C.

Using iDPC STEM, the projected polarization is measured across ten different sample locations using the approach outlined in Supplementary Information Section \ref{supplsec:projectedPolarization}.  This yields approximately 5,000 analyzed unit cells at each of the three PMN-PT compositions.  The high sampling of each composition provides a statistical representation of the underlying structure of the material, aiding in connecting the results presented here to prior diffraction studies. Calculating the center of mass difference between the cations and anions using iDPC STEM, Figure \ref{fig:quiver}c shows representative projected polarization maps for PMN, PMN-10PT, and PMN-30PT imaged along \oneoneo{}. The projected polarization exhibits regions with similar magnitude and direction, i.e.~polar domains, that vary smoothly across each image.

Notably for PMN, the observed polar domains do not decay to a non-polar background, which contrasts with the 'polar nanoregion' model that assumes PNRs exist in a non-polar matrix. These nanoscale domains vary in size from 2-12 nm and form low angle domain walls between them, which are located using the method described in Supplementary Information Section \ref{supple:domainwalldistribution}.  This observation is also remarkably similar to reverse Monte Carlo analysis of diffraction data.\cite{Eremenko2019}.  Moreover, a significant majority (72\%) of the domain wall angles are in the range of 10-35$^\circ$ , which is in agreement with predictions from the `polar slush' model (see Supplementary Information Section \ref{supple:domainwalldistribution}).\cite{Takenaka2017,Kim2019} As the fraction of PT increases, the average domain wall angle increases and  reflects the onset of ferroelectric behavior. This behavior leads to the mixed relaxor and ferroelectric properties found in PMN-PT materials and expected from the polar slush model.\cite{Takenaka2017,Kim2019} The polar slush model does not, however, incorporate the specific structure and chemistry details that drive the formation of the polar domains.   

% It is also observed that the domain size does not vary with decrease in temperature rather the correlation inside a domain increases which agrees with our model as structural and chemical heterogeneities does not vary with decrease in temperature leading to no change in the domain size. 

%Domain walls occur at regions where the projected polarization has the greatest local variation on the scale of 1-2 unit cells.  Thus, domain walls will occur at positions where the local standard deviation (Supplementary Information section 1E) is high. 

\begin{figure}
    \centering
    \includegraphics[width = 6.4 in]{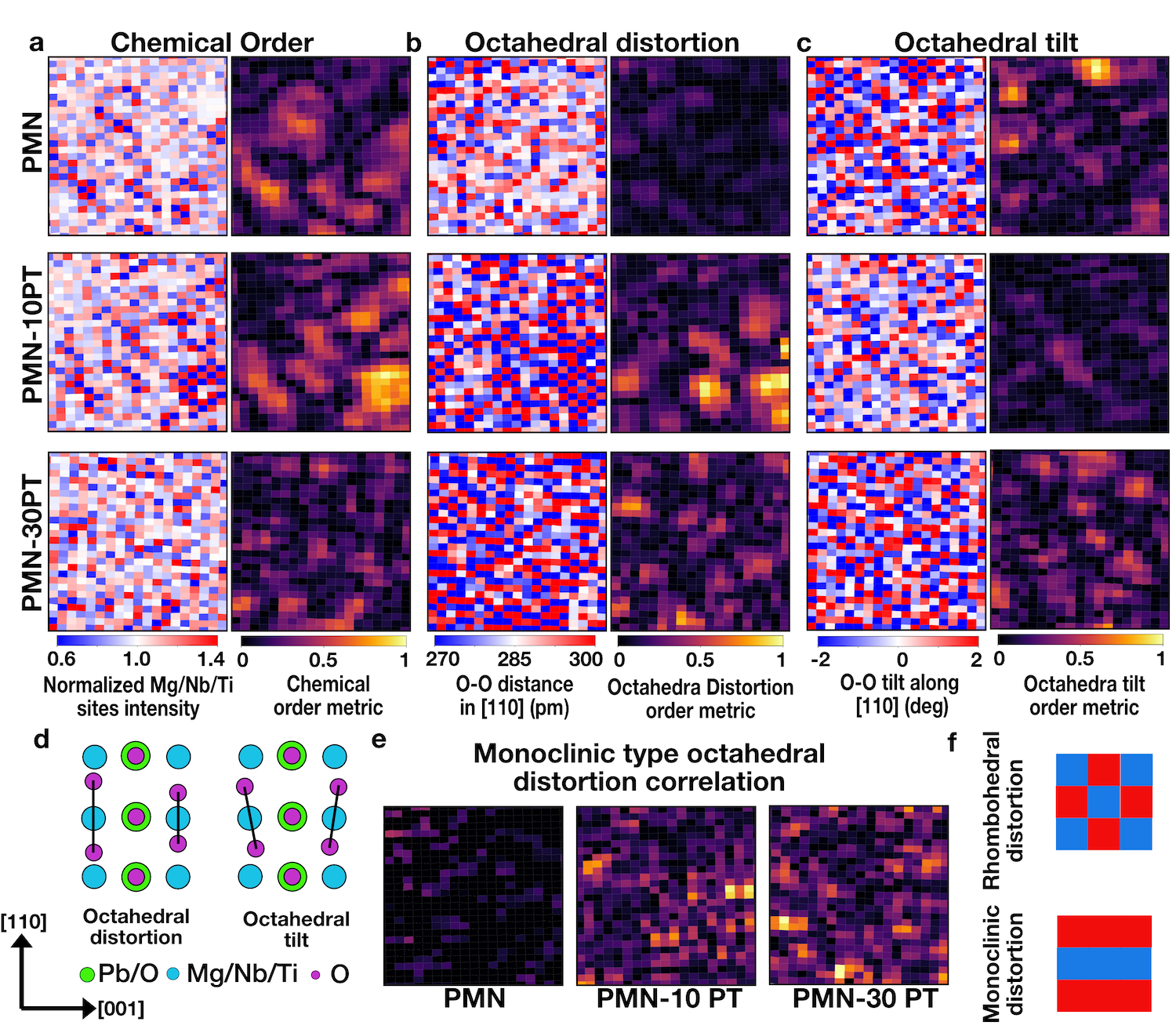}
    \caption{Spatial distribution of structural and chemical heterogeneities, (a) chemical order, (b) octahedral distortion, (c) octahedral tilt, (d) schematic of distortion types and their corresponding O-O patterns, (e) monoclinic-type distortion analysis, and (f) schematic patterns of rhombohedral and monoclinic distortion.}
    \label{fig:Inhomo}
\end{figure}

Chemical ordering is revealed in the ADF STEM data, as shown in Figure \ref{fig:Inhomo}a,left. Clustered oscillation of B sub-lattice atom column intensity on (111) planes indicates a doubled perovskite lattice where occupancy of Mg and Nb vary systematically. The weaker normalized intensity atom columns (blue) contain more Mg or Ti ($\beta_I$), while those stronger intensity atom columns (red) contain more Nb ($\beta_{II}$). Correlation analysis, shown in Figure \ref{fig:Inhomo}a and detailed in Supplementary Information Section \ref{supplsec:correlationAnalysis}, is used to quantify the relative fraction of these ordered regions. At each composition, the CORs are found throughout and account for 39$\pm$1\% of the total projected area in the case of PMN,  37$\pm$2\% for 10\% of PT, and only 11$\pm$1\% for 30\% PT.  The decreasing chemical order with increasing Ti is in agreement with previous X-ray and neutron scattering results where superlattice reflection intensity also decreases as the composition approaches the MPB.\cite{Hilton1990}

Recently, Krogstad suggested the presence of anti-ferrodistortive displacements based on diffuse scattering features, i.e.~Pb displaces in the opposite direction of its B sub-lattice neighbors \cite{Krogstad2018}.  The chemical and spatial origins of this behavior could not, however, be unambiguously determined.  From iDPC STEM data, the Pb atom columns are anti-ferrodistortive relative to the B-sites largely in the CORs, (Supplementary Information Section \ref{supple:pbDisplacement}). These anti-ferrodistortive displacements decrease in accordance with the decreasing COR density as  PT content increases, and in agreement with Krogstad.

% Furthermore, the regions between the CORs are the first to exhibit enhanced polarization The addition of Ti in combination with the polarization from iDPC also reveals that , see Supplementary Information SX. 

As noted from recent X-ray and neutron scattering experiments, oxygen displacements may hold the key to understanding structure-property relationships in Pb-based relaxors.\cite{Krogstad2018} To this end, oxygen octahedral distortion (expansion/compression) and tilting (schematically shown in Figure \ref{fig:Inhomo}d) are determined from iDPC images, as in Figures \ref{fig:Inhomo}b-c. At each composition, distortion and tilting exhibit local ordering, reminiscent of the CORs.

% The  checkerboard pattern corresponding to  

\begin{figure}
    \centering
    \includegraphics[width = 3.2in]{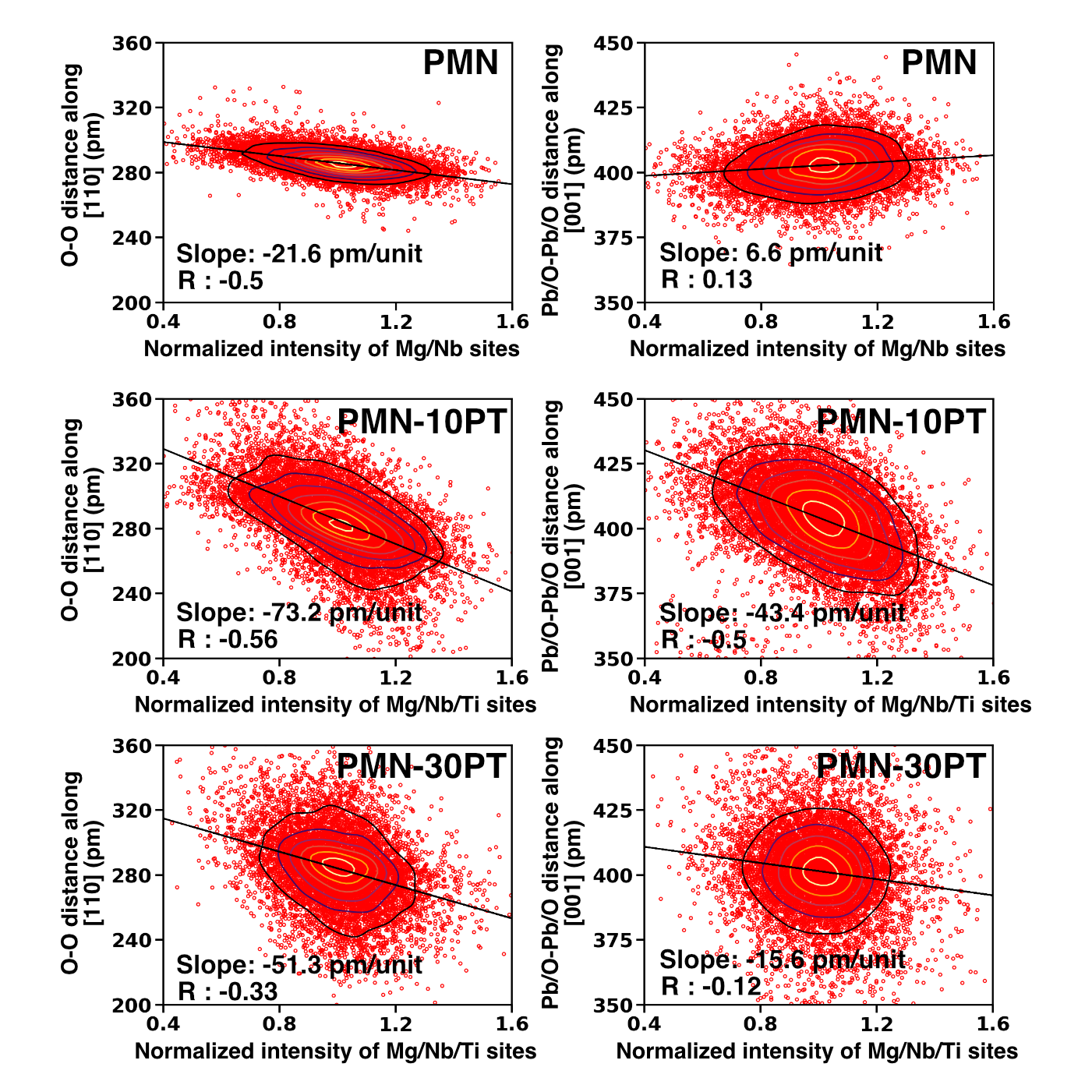}
    \caption{Correlation between the Mg/Nb/Ti normalized atom column intensity and the O-O distance along \oneoneo{} or Pb/O-Pb/O distance along $\left[001\right]$ as indicated.}
    \label{fig:OODistCor}
\end{figure}

Correlation analysis is used to determine the area fraction of octahedral distortion regions (ODRs), as shown in Figure \ref{fig:Inhomo}b. In PMN, 21\% of the projected area exhibits distortion ordering, with 63\% overlapping the CORs and 27\% at the COR boundaries. As PT content increases, the ODRs increase to 28\% for PMN-10PT. Furthermore, a similar checkerboard octahedral distortion pattern is predicted for the ordered PMN structure from density functional theory (DFT) (Supplementary information Sections \ref{supple:dft} and \ref{supple:stemImageSims}).

At PMN-30PT, however, the distortion order changes.  Rather than exhibiting the rhombohedral-type, checkerboard   distortion pattern (see Figure \ref{fig:Inhomo}f), the distortion becomes predominately striped on $\left(110\right)$ planes, which indicates the formation of monoclinic-like unit cells as verified with STEM image simulations (Supplementary Information Figure \ref{fig:STEM_crystalSIm}) and agrees with prior diffraction studies \cite{Singh2006}. Analyzed across the composition space,  monoclinic-type distortion increases with addition of Ti as shown in Figure \ref{fig:Inhomo}b. Moreover, this type of local planar distortion should give rise to asymmetric diffuse scattering \cite{Kreisel2003}.  Importantly, Krogstad \cite{Krogstad2018} found that the oxygen mediated asymmetric diffuse scattering in neutron scattering was the only feature linked to the piezoelectric properties. The results presented here confirm that oxygen displacements are key and they originate in monoclinic-type oxygen octahedral regions.

Beyond expansion and contraction, the oxygen octahedra also tilt with respect to \oneoneo{}, as shown in Figure \ref{fig:Inhomo}c. Applying correlation analysis, 19\% of the total projected area exhibits octahedral tilt ordering in PMN. Furthermore, the B sub-lattice atom column intensities in octahedral tilt regions (OTRs) are normally distributed (see Supplementary Information Section \ref{supple:intensityDistribution}). This reveals that the OTRs form in the regions of randomly distributed Mg and Nb instead of the CORs. The anti-phase tilt pattern suggest the presence of local rhombohedral, $R3c$, or orthorhombic, $Pnma$ local symmetry.\cite{Glazer1972} With increasing PT, the OTR area decreases from 11\% to 5\% for 10PT and 30PT, respectively.  As the fraction of these features decrease with composition, their disappearance also explains another distinct diffuse scattering contribution noted by Krogstad that had both temperature dependent and independent components. While the temperature independent contribution is thought to originate from the CORs, the temperature dependent part then arises from the octahedral tilt ordering that is  disrupted by thermal fluctuations. The intensity of this type of diffuse scattering is found to decrease with PT content and follows the same trend as OTRs quantified from STEM.

%importance of oxygen to this contribution as X-ray scattering is dominated by the cations while neutron scattering is sensitive to both the cations and anions. Thus, it indicates that C1 intensity arises from these monoclinic type distortion in oxygen sublattice as observed in STEM analysis.

Combining the ADF and iDPC STEM data, the change in B site chemistry measured with the Mg/Nb/Ti atom column intensity is also found to correlate to the O-O neighbor distance, as shown in Figure \ref{fig:OODistCor}.  In each case, the B sub-lattice intensity is negatively correlated with the O-O atom column distances \cite{Sang2015}. This shows that Mg/Ti expands the local oxygen sub-lattice, while Nb leads to contraction. In combination with the response of oxygen at the MPB where the O-O spread increases, the observed correlation indicates the key role of Ti to disrupt the lattice to stabilize the monoclinic phase and increase the piezoelectric coefficient. The Pb/O-Pb/O atom column distances, in comparison, show weak-to-no correlation with variation in B-site chemistry in PMN as shown in Figure \ref{fig:OODistCor}, likely due to the dominant contribution of Pb on the observed atom column positions (Supplementary Information Section \ref{supplsec:projectedPolarization}). In contrast, PMN-10PT exhibits moderate, negative correlation.  At the MPB, PMN-30PT, significantly larger distortions are observed that correspond to the increased polarization and piezoelectric coefficient. This correlated structural variation in Pb displacement also explains the diffuse scattering feature found by Krogstad that correlates to the cation size mismatch.

Further analysis of the PMN-10PT $\beta_{I}$ and $\beta_{II}$   normalized intensities (Supplementary Information Figure \ref{supple:intensityDistribution}) also reveals  that their intensity difference is enhanced with the introduction of Ti. This suggests that Ti initially replaces Nb in the mixed $\beta_I$ sites of the CORs leading to a concomitant decrease in intensity for those atom columns and increased image contrast. Importantly, the preferential segregation of Ti to the CORs has not been previously shown and can be explained by analysis of the B sub-lattice bond lengths from DFT. Any covalent bond has an ideal length or range of lengths it prefers to adopt for any particular coordination environment. For example in oxide perovskites, titanium tends to prefer bond lengths between 1.9 and 2.1 $\text{\AA}$ \cite{abramov1995chemical, cole1937lead, shin2007order, yoshiasa2016high}. From DFT, Supplementary Figure \ref{fig:Bsites-within-constraints.png}, the COR structure has more sites close to Ti's ideal octahedral bonding environment than the disordered structure. These results indicate that Ti can form more, strong bonds in the COR structure than the disordered structure, without changing any other bond lengths or otherwise altering the structures. All else being equal, this strongly suggests that Ti will preferentially incorporate  into the CORs, at least on the PMN side of the composition space where CORs form.

\begin{figure}
    \centering
    \includegraphics[width = 3.2in]{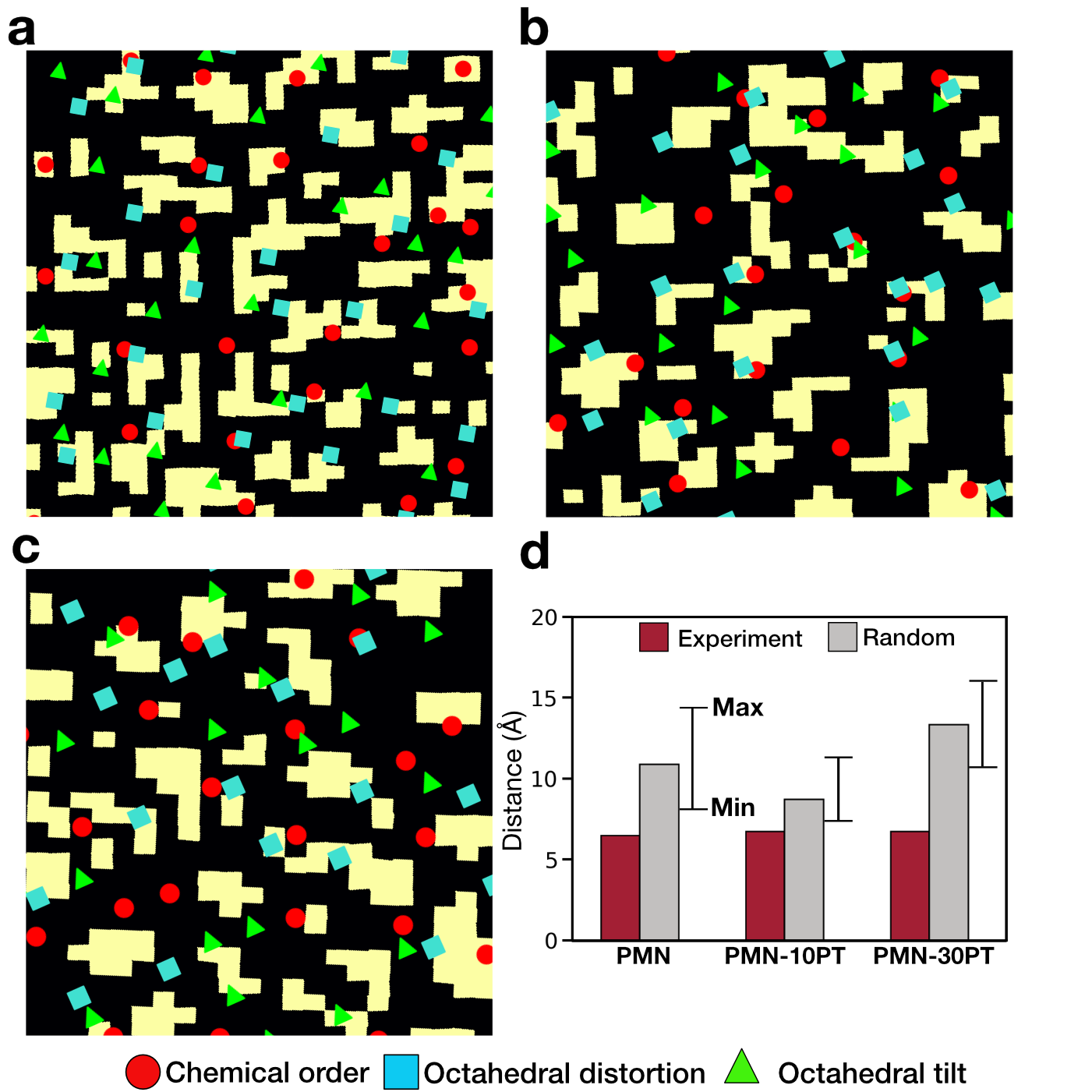}
    \caption{Positions of CORs, ODRs \& OTRs and domain walls for (a) PMN, (b) PMN-10PT, and (c) PMN-30PT. (d) Nearest distance between 95 \% of  heterogeneities maxima (random points) and domain wall for each PMN-xPT composition. Domain walls occur at the blocked, yellow regions.}
    \label{fig:corPolInhomo}
\end{figure}

The heterogeneity distribution is also found to link directly with the local variation in polarization and domain walls. First, the centers of the CORs, ODRs and OTRs are identified as maxima  in the results from correlation analysis as detailed in Supplementary Information Section \ref{supple:polarCorrelation}.  Qualitatively, Figure \ref{fig:corPolInhomo}a-c, the heterogeneity maxima occur at, or very near, the locations of greatest polarization variation, i.e.~low angle domain walls. 

The observed relationship between ordered heterogeneities and domains walls is validated by quantifying the nearest neighbor distance. As shown in Figure \ref{fig:corPolInhomo}d, 95$\%$ of the heterogeneities are within a distance of 1.5-2 unit cells of the nearest domain wall regardless of PT content. Across the composition range investigated, the 1/3 of CORs, 1/3 ODRs, and 1/3 OTRs within this distance of the domain walls. In addition, 1/3 of the domain walls have at least two nearby types of structural or chemical ordering. This indicate that the polar domain walls cannot be explained by a single type of heterogeneity, but requires the consideration of all features identified.

To determine the significance of this observation, consider the null-hypothesis that heterogeneities are randomly distributed with respect to polarization.  Taking 10,000 random sets of data equivalent to experiment, but with random heterogeneity locations, the 95$\%$ threshold is not reached until a distance two times larger than experiment, see Figure \ref{fig:corPolInhomo}d.  Further highlighting the difference, the error bars indicate the maximum and minimum 95$\%$ distances for the 10,000 datasets at each composition. A result similar to experiment is not found any of the random data, indicating that $p$ is significantly less than 0.05, safely discounting the null-hypothesis. Taken together, the results presented here unambiguously reveal that chemical and structural heterogeneities are not randomly distributed with respect to polarization, but instead act as the mechanism to disrupt the formation of long range polarization. Although this role heterogeneity has been postulated, its connection to generating low angle domain walls has not been directly observed until now.\cite{Randall1990, Fu2009}

In conclusion, direct quantification of cations and anion atom columns from  STEM has revealed  key relationships between polarization, structure, and chemistry in the prototypical relaxor ferroelectric system, PMN-PT. The presence of nanoscale domains exhibiting low angle domain walls is in agreement with the recently proposed polar slush model\cite{Takenaka2017}. The formation of nanoscale polar domain walls arises from the presence of short range chemical and/or structural heterogeneity, which act to disrupt the long range polarization.  The introduction of Ti decreases the type and number of heterogeneities, which leads to cooperative alignment of polarization while also increasing the fraction of monoclinic distortion that enhances the piezoelectric coefficient. The  combination of nanoscale features explains the origin for relaxor ferroelectric behavior in PMN-PT that lead to the dramatically improved piezoelectric performance.  Extending this knowledge, it is anticipated that engineering a combination of short range structural and chemical order are required for advancing the next generation of Pb-free relaxor-ferroelectric materials. 

%This opens the path to designing new relaxor-ferroelectric materials by searching for additions that control heterogeneities in the system.

%amount of various heterogeneities in materials will lead to develop better engineered ferrolectric/piezoelectric materials. 

% \begin{figure}
%     \centering
%     \includegraphics[width = 6in]{Images/Fig1.png}
%     \caption{(a)Displacement of center of mass of cations from anions. (b) Displacement of A site with respect to B site.}
%     \label{fig:Quiver}
% \end{figure}

% \begin{figure}
%     \centering
%     \includegraphics[width = 6in]{Images/Fig2.png}
%     \caption{(a) B site ordering of beta I and beta II atom coulmns. (b) Order metric calculated for B-site intensities.}
%     \label{fig:OMetric}
% \end{figure}

% \textbf{Statistical analysis}\\

% %A similar correlation of B site chemistry with structural changes in oxygen is found across images from 10 different locations in the sample (correlation coefficient: 0.4-0.71).

\newpage

\bibliographystyle{Science}
\bibliography{refs}

\newpage

% \begin{figure}
%     \centering
%     \includegraphics[width = 3.3in]{Images/Quiver.png}
%     \caption{Cation displacement with respect to anion positions.}
%     \label{fig:Quiv}
% \end{figure}

% \begin{figure}
%     \centering
%     \includegraphics[width = 6in]{Images/ComparisonOMtilt.png}
%     \caption{(a) Displacement direction with for angles in 8 bins. (b) Order metric calculated for B-site intensities. (c) O-O distance distribution.}
%     \label{fig:OMcomp}
% \end{figure}

\section*{Acknowledgements}

We gratefully acknowledge support for this work from the National Science Foundation, as part of the Center for Dielectrics and Piezoelectrics under Grant Nos.~IIP-1841453 and IIP-1841466. SZ acknowledges support from the Australian Research Council (FT140100698) and the Office of Naval Research Global (N62909-18-12168).  PCB was supported by the Department of Defense through the National Defense Science \& Engineering Graduate (NDSEG) fellowship program.
Computational time and financial support for JNB was provided by AFOSR grant FA9550-17-1-0318.
MJC acknowledges support from National Science Foundation as part of the NRT-SEAS under Grant No.~DGE-1633587. This work was performed in part at the Analytical Instrumentation Facility (AIF) at North Carolina State University, which is supported by the State of North Carolina and the National Science Foundation (ECCS-1542015). AIF is a member of the North Carolina Research Triangle Nanotechnology Network (RTNN), a site in the National Nanotechnology Coordinated Infrastructure (NNCI). We acknowledge Matthew Hauwiller for useful suggestions while preparing the manuscript.

\section*{Author Contributions}

AK conducted the electron microscopy experiments, data analysis, and image simulations. MC prepared the PMN samples for electron microscopy and collected STEM data. SZ grew the PMN-xPT single crystals. JB and DL performed the DFT calculations and the corresponding analysis.  JML and ECD designed the electron microscopy experiments and guided the research.  All authors co-wrote and edited the manuscript.

\section*{Competing interests}
The authors declare no competing interests.

\section*{Methods}
\subsection*{Sample Information}

Pb(Mg$_{1/3}$Nb$_{2/3}$)O$_{3}$-xPbTiO3 (PMN-xPT) single crystals were grown via the high temperature flux method \cite{Zhang2012}. Samples for electron microscopy were cut from these larger single crystals and oriented along $\left< 011\right>$ and then thinned to electron transparency using mechanical wedge polishing \cite{Voyles2002a} followed by low temperature, low energy Ar ion milling.  

\subsection*{Scanning Transmission Electron Microscopy}

STEM imaging was performed with a probe-corrected FEI Titan G2 60-300 kV S/TEM equipped with an X-FEG source at a beam current of 30 pA and probe semi-convergence angle of 19.6 mrad.  We used a custom scripting interface to automate the Thermo Fisher Scientific Velox software for simultaneous ADF and iDPC acquisition.  ADF images were collected with a semi-collection angle range of 34-205 mrad and iDPC with a with collection semi-angle range of 7-28 mrad. For high image accuracy and precision, the revolving STEM (RevSTEM) method was used.\cite{Sang2014c,Dycus2015} Each RevSTEM data set consisted of 20 1024$\times$1024 pixel frames with a 90$^\circ$ rotation between each successive frame. Sample thickness ranged from 6 to 10 nm as determined using position averaged convergent beam electron diffraction \cite{Lebeau2010}. Supercells were constructed using relaxed structures from DFT, see next section. A repeating unit was cropped and simulated using a custom python based STEM image simulation software. Simulated ADF \& iDPC images were convolved with a Gaussian with full-width at half-maximum of 80 pm to approximately account for the finite effective source size \cite{Lebeau2008}.

\subsection*{Density Functional Theory and Image Simulations}

Density functional theory (DFT) calculations of PMN were performed to investigate the atomic structures in three dimensions, and to examine charge localization and bond length distributions.
These calculations were performed with the PBE exchange correlation functional in VASP 5.3.3, collinear spin polarization, and a plane wave kinetic energy cutoff of 520 eV. \cite{Kresse1993,Kresse1994,kresse1996efficiency,Kresse1996a}
A single reciprocal space point at $\Gamma$ was used. 
Projector Augmented Wave pseudopotentials were used with 2, 11, 4, and 6 valence electrons explicitly treated for Mg, Nb, Pb, and O, respectively.  Additional details of the calculation are provided in Supplementary Information Section \ref{supple:dft}.

\end{document}